\documentclass[aps,prd,twocolumn,epsf,showpacs,superscriptaddress]{revtex4}

\usepackage{graphicx}
\usepackage{amsmath,amssymb,latexsym}
\usepackage{bm} 

\def\Tr{\text{Tr}} 

\begin{document}

\title{Detecting monopoles on the lattice}

\author{Claudio Bonati}
\affiliation{Dipartimento di Fisica, Universit\`a
di Pisa and INFN, Sezione di Pisa, Largo Pontecorvo 3, 56127 Pisa, Italy}
\author{Adriano Di Giacomo}
\affiliation{Dipartimento di Fisica, Universit\`a
di Pisa and INFN, Sezione di Pisa, Largo Pontecorvo 3, 56127 Pisa, Italy}
\author{Massimo D'Elia}
\affiliation{Dipartimento di Fisica, Universit\`a
di Genova and INFN, Sezione di Genova, Via Dodecaneso 33, 16146 Genova, Italy\\}

\begin{abstract}
We address the issue why the number and the location of magnetic monopoles detected on lattice configurations
are gauge dependent, in contrast with the physical expectation that monopoles have a gauge-invariant status. By use
of the non-Abelian Bianchi identities we show that monopoles are gauge-invariant, but the efficiency of the technique 
usually adopted to detect them depends on the choice of the gauge in a well understood way. In particular we have studied 
a class of gauges which interpolate between the Maximal Abelian gauge, where all monopoles are observed, and the Landau 
gauge, where all monopoles escape detection.
\end{abstract}

\maketitle

\section{Introduction}
Monopoles may play a fundamental role in gauge theories, their condensation in the vacuum being
responsible for dual superconductivity and color confinement\cite{'tHP,m,'tH2}.

The correct way to detect monopole condensation is to measure the vacuum expectation value of an operator $\mu$ carrying 
nonzero magnetic charge: if $\langle \mu\rangle \neq 0$ vacuum is a dual superconductor 
\cite{digz}. In the deconfined phase $\langle \mu\rangle=0$. Much work has been done on this line 
\cite{dp,ddp,ddpp,dlmp,ccos,3,vpc}.

Much more activity, however, has been devoted to the observation of monopoles in lattice configurations, either 
to look for an effective monopole-action, based on the empirical observation that monopoles could be the dominant
degrees of freedom of the theory (monopole dominance)~\cite{sco,suz,pol}, or, more recently, to study the relevance of 
thermal magnetic monopoles to the properties of the deconfined phase~\cite{zak,bornya,ejiri,del,chern,del1}.

The detection of monopoles in lattice configurations is a highly non trivial problem. The procedure is well defined and 
gauge invariant in compact $U(1)$ gauge theory \cite{dgt}: any excess over $2\pi$ of the Abelian phase of a plaquette is 
interpreted as existence of a Dirac string through the plaquette; a net magnetic charge  exists in an elementary cube when a 
net number of Dirac  flux lines  crosses the plaquettes  at its border. In this model the phase of the elementary Wilson 
loop is gauge invariant and therefore the procedure is unambiguous. In the case of a non-Abelian gauge theory, instead, one 
has first to fix a gauge, and then to apply the same procedure to the Abelian subgroup spanned by some diagonal component of 
the Lie algebra \cite{'tH2}. The result strongly depends on the choice of the gauge and, as a result, the existence of a 
monopole in a location of a given lattice configuration is a gauge dependent property, and this is of course physically 
unacceptable. 
 
In the soliton monopole of Refs.~\cite{'tH,Pol} the Abelian subgroup which identifies the magnetic $U(1)$ coupled to the 
magnetic charge coincides with the invariance subgroup of the vacuum expectation value of the Higgs field which produces the 
symmetry breaking. In QCD there is no Higgs field, but the magnetic $U(1)$ has to be a subgroup of the gauge group, due to 
the general property that the monopole component of any field configuration is intrinsically Abelian \cite{coleman}.
It was proposed in Ref.~\cite{'tH2} that any operator in the adjoint representation could be used as an effective Higgs 
field to identify the magnetic $U(1)$ subgroup, the physics being for some reason independent of that choice. Each choice 
was called an Abelian projection. In practice different  Abelian projections proved to have different features, and there 
was a general consensus on the choice of the so-called ``maximal Abelian gauge''~\cite{sch} as the most convenient to 
expose Abelian dominance and monopole dominance~\cite{suz,pol}. More recent studies have shown that the differences between 
Abelian projections are less important than they look in simulations with low statistics~\cite{sco,scol}, but the question 
of understanding the differences between Abelian projections is still important.
  
In a previous paper~\cite{bdlp}, which we shall quote as I in the following, these problems have been analyzed with the 
following results:
\begin{enumerate}
\item The magnetic currents in any Abelian projection are proportional to the violation of the non-Abelian Bianchi 
identities, which is gauge covariant. They are the projection of that violation on the fundamental weights which 
identify the Abelian projections.
  
\item Magnetic currents observed in lattice configurations depend on the Abelian projection.
  
\item  For each field configuration there exists a preferred direction in color space, which is that of the (Abelian)
magnetic monopole term in the multipole expansion~\cite{coleman}. That direction is the same that is identified by the 
maximal Abelian gauge. 
  
\item  Only in the maximal Abelian gauge (modulo an arbitrary global transformation times a gauge transformation which is 
trivial at infinity) does the magnetic charge obey the Dirac quantization condition. In other gauges the charge measured by 
the flux of the magnetic field at large distances is generically smaller than the true monopole charge.
  
\item Monopole condensation is a gauge-invariant feature.
\end{enumerate}
  
Since the recipe of Ref.~\cite{dgt} for counting monopoles is  based on the Dirac quantization condition, all this is 
equivalent to state that the existence of a monopole is a gauge-invariant fact, but the detection recipe is Abelian 
projection dependent.
  
All these findings were analytic and were analytically tested on the soliton solution of Refs.~\cite{'tH,Pol}.
  
In the present paper we want to check them numerically on Monte Carlo generated field configurations. As shown in 
Refs.~\cite{bdlp,DLP}, proving this for $SU(2)$ gauge theory in absence of fermions is equivalent to proving it for the 
generic case. We shall therefore simulate quenched $SU(2)$.
  
In Sec. II we show  that the unitary gauge is nothing but the maximal Abelian gauge, as already stated in 
Ref.~\cite{bdlp}.  We  then note that the ``hedgehog'' gauge of Refs.~\cite{'tH,Pol} is nothing but the Landau gauge, and 
that in this gauge the flux at infinity of the magnetic field as defined by the 't Hooft tensor is vanishing. This explains 
why no monopole is found in the Landau gauge by use of the recipe of Ref.~\cite{dgt} (see \emph{e.g.} \cite{suzu}).
  
We  then introduce a class of gauge transformations continuously dependent on one parameter, connecting these two gauges, and 
we compute the expectation of the observed number of monopoles along this connection, to be compared with the numerical 
data. The idea is that in the maximal Abelian gauge all the monopoles are detected, while, by changing the parameter of 
the gauge transformation, the effective magnetic charge is decreased and so is the number of observed monopoles, which is 
finally zero in the Landau gauge.
   
Sec. III reports the numerical simulations and the result of the check .
   
In Sec. IV we  discuss the results and we conclude.

\section{From maximal Abelian to Landau gauge}

We consider the soliton solution of Refs.~\cite{'tH,Pol}. It was shown in the paper I that the result also applies to
a generic configuration. 
With the notation of Ref.~\cite{shnir} the gauge field of the soliton solution in the hedgehog gauge is
\begin{equation}
\begin{array}{l}
A^a_0 = 0 \\ A^a_n = - (1-K(r))\frac{\displaystyle \epsilon_{a n j} r^j}{\displaystyle gr^2} 
\end{array}
\label{tph}
\end{equation}
where $a$ is the color index, $(0,n)$ the space time indexes, and $K(r)$ a form factor vanishing for large $r$ whose 
specific form depends on the parameters of the Higgs sector. It is trivial to check that the solution in \eqref{tph} obeys 
the Landau gauge condition
\begin{equation}
\partial_{\mu}A^a_{\mu} =0
\end{equation}
The 't Hooft tensor in this gauge is by definition
\begin{equation}
F_{\mu \nu} = \partial_{\mu} A^3_{\nu} - \partial_{\nu} A^3_{\mu}
\end{equation}
and, by using the expressions in Eq.~(\ref{tph}) and recalling that the Abelian fields are given by $e_i=F_{0i}$ and 
$b_i = {1\over 2}\epsilon_{ijk} F_{jk}$, respectively, one gets for large $r$
\begin{eqnarray}
\vec e=0, \qquad b_i = -2\frac{r^i z}{g r^4} 
\end{eqnarray}
The magnetic charge $Q$ which is detected by the recipe of Ref.~\cite{dgt} is proportional to the flux of $\vec b$ across a 
spherical surface at $r \to \infty$, so one gets
\begin{equation}
Q_m = \int \mathrm{d}\Omega\, (\vec b \cdot \vec n) = {2 \over g} \int \mathrm{d}\Omega\, \cos\theta=0
\end{equation}
No magnetic charge is thus expected in the Landau gauge.
  
In the unitary gauge, instead, the field obeys the gauge condition which defines the maximal Abelian gauge, namely,
\begin{equation}
\partial_{\mu} A^{\pm}_{\mu} + ig \left[A^3_{\mu}, A^{\pm}_{\mu}\right] = 0 \label{mag}
\end{equation}
This can be easily checked by direct computation on the explicit solution given in the appendix of Ref.~\cite{shnir}. The exact 
field at all distances $ \bar A_{\mu}$ in the unitary gauge, which is given in Eq.~\eqref{maf} below, exactly obeys 
Eq.~\eqref{mag}:
\begin{eqnarray}
\bar A_0  &=&0 \nonumber \\
\overrightarrow{\bar{A}\,\,} &=& -\frac{1}{2gr} \Bigg\{ \hat \phi \left( \frac{\cos\Theta -1}{\sin\theta} + (1-K)\sin(\Theta - \theta)\right)
\sigma_3 \nonumber \\ 
&+& \Bigg[ \hat \phi \left((1-K) \cos(\Theta - \theta) -\frac{\sin\Theta}{\sin\theta}\right) \sigma_1 + \nonumber \\
&+& \hat \theta (\Theta' -1+K)\sigma_2 \Bigg] (\cos(\phi) +i\sigma_3 \sin(\phi) ) \Bigg\} \label{maf}
\end{eqnarray}
Here $\Theta =\theta\frac{1+\cos\theta}{1+\cos\theta +\epsilon^2}$ is a regulator of the singularity at $\theta =\pi$ and 
$\Theta'$ its derivative with respect to $\theta$; $\hat{\phi}$ and $\hat{\theta}$ are the versors of the $\phi$ and $\theta$ 
directions.

As shown in the paper I in this Abelian projection the Abelian magnetic charge is equal to two units, and can be detected 
by the recipe of Ref.~\cite{dgt} since the Dirac condition is satisfied. The unitary transformation from the unitary (maximal 
Abelian) gauge to the hedgehog (Landau) gauge is known to be
\begin{equation}
\begin{array}{c}
A_{\mu}= U \bar{A}_{\mu} U^{\dag}+\frac{i}{g}(\partial_{\mu}U)U^{\dag} \\
\rule{0mm}{5mm}U = \exp\left(-i\phi {\sigma_3 \over 2}\right) \exp\left(-i\Theta { \sigma_2 \over 2}\right) 
\exp\left(i\phi {\sigma_3 \over 2}\right) 
\end{array}
\label{gtr}
\end{equation}
We shall operate a class of gauge transformations $U(\alpha)$ on the configurations in the maximal Abelian gauge, depending 
on one parameter $\alpha$, such that for $\alpha=0$ we stay in the maximal Abelian gauge [$U(\alpha)=1$] while for 
$\alpha =1$ we have $U(\alpha)=U$ of Eq.~\eqref{gtr} and we go to the Landau gauge. For each value of the parameter 
$\alpha$, $U(\alpha)$ identifies an Abelian projection and the corresponding Abelian magnetic charge $Q_m(\alpha)$ can be 
computed. We already know that $Q_m(0)=2$ and $Q_m(1)=0$. The general form of this transformation is
\begin{eqnarray}
U(\alpha) = \exp\left(-i\gamma(\theta, \phi, \alpha) {\sigma_3 \over 2}\right) \times \hspace{2.5cm}\nonumber\\
\times \exp\left(-i\beta(\theta,\phi,\alpha){ \sigma_2 \over 2}\right) 
\exp\left(i\gamma(\theta, \phi, \alpha){\sigma_3 \over 2}\right) \label{fma}
\end{eqnarray}
with $\beta, \gamma$ functions of the polar angles $\theta, \phi$ and of the parameter $\alpha$, with boundary conditions 
\begin{eqnarray}
\gamma(\theta,\phi,1)=\phi \quad \beta(\theta,\phi,0)=0 \quad \beta(\theta,\phi,1)=\Theta
\end{eqnarray}
We shall work out in detail the special case
\begin{equation}
\gamma(\theta,\phi,\alpha)= \phi \quad \beta (\theta,\phi,\alpha)= \alpha \Theta 
\end{equation}
 
We are interested in the radial component of the Abelian magnetic field at large distances and along $\sigma_3$ in color 
space. At large distances and $\theta \neq \pi$ Eq.\eqref{maf} gives for the field
\begin{equation}
\vec A = - \frac{1}{2gr} \hat \phi\sigma_3\frac{\cos \theta -1}{\sin \theta}
\end{equation}
Operating the transformation \eqref{fma} and projecting on the third axis gives
\begin{equation}
\vec  A^{\alpha}_3 \equiv \Tr \left[ \frac{\sigma_3}{2} \vec{A}^{\alpha} \right] = 
\frac{1 - \cos \theta \cos(\alpha \theta)}{2gr\sin \theta}\hat{\phi}
\end{equation}
and for the Abelian magnetic field 
\begin{eqnarray}
\vec b(\alpha) = \hat r \frac{1}{r \sin \theta}\partial_{\theta}(\sin\theta \, \hat{\phi}\cdot\vec{A}^{\alpha}_3)
\end{eqnarray}
For the  magnetic flux at infinity we have 
\begin{equation}
\Phi (\alpha)= r^2 \int \mathrm{d}\Omega\, \hat{r} \cdot \vec{b}(\alpha) = \frac{\pi}{g} \big[1 + \cos(\alpha \pi)\big]
\end{equation}
and for the ratio of the magnetic charge to that of the maximal Abelian projection we finally get
\begin{equation}
\frac{Q_m(\alpha)}{Q_m(0)} = \frac{1 + \cos(\pi \alpha)}{2} \label{qratio}
\end{equation}
This result is exact for the 't Hooft-Polyakov monopole and, by use of the same argument introduced in the paper I, it is 
also exact for a generic monopole in the continuum theory. 

On the lattice Eq.~\eqref{qratio} should give the ratio of the number of monopoles observed in the gauge corresponding to a 
generic value of $\alpha$ to that observed in the maximal Abelian gauge. Of course on the lattice there are discretization errors
which are not taken into account in the derivation of Eq.~\eqref{qratio}: a lattice configuration in the maximal Abelian gauge 
satisfies Eq.~\eqref{mag} only approximately, the position of the monopole is not determined with arbitrary precision, the string direction is    
known with an angular error of $\pi/4$, and the gauge field is constant along links.

Because of these discretization errors we can not expect Eq.~\eqref{qratio} to be exactly satisfied in lattice simulation. Nevertheless,
we expect it to hold qualitatively: by increasing the parameter $\alpha$ from $0$ to $1$ 
in a given configuration no new monopole should appear, but some of them should disappear, till the Landau gauge is reached at $\alpha=1$, where 
no monopole should be observed.

\section{Numerical results}

As anticipated in Sec. I we shall test the above analysis in quenched $SU(2)$ since the results can trivially 
be exported to a generic gauge theory with and without quarks. 

To simulate $SU(2)$ gauge theory we used a standard combination of heatbath \cite{creutz80,kp} and overrelaxation 
\cite{creutz87} updates. 
The maximal Abelian gauge fixing was achieved by an iterative combination of local maximization and overrelaxation steps 
(see \emph{e.g.} the appendix of Ref.~\cite{cc}) and the algorithm was stopped when the average square modulus of the non 
diagonal part of the operator $X(r)$, which has to be diagonal in the maximal Abelian gauge, was less than $10^{-11}$. 
Abelian links are then extracted as the third component of the gauge fixed field and monopoles are located by using the 
recipe of Ref.~\cite{dgt}.

On this last point some care has to be used; for Eq. \eqref{qratio} to be true it is mandatory to integrate the 
Abelian magnetic flux on a surface at infinity, otherwise a residual dependence on the function $K$ in Eq.~\eqref{tph},
\eqref{maf} is present. On the lattice monopoles are usually detected by integrating the flux on a cube whose faces are the 
elementary plaquettes. Is this sufficient to obtain the charges? It is possible to answer this 
question by using larger cubes, whose faces are $n\times n$ Wilson loops with $n>1$: if the result is independent of
$n$ this means that elementary cubes are sufficient. The observed behavior is different in different gauges: while for the
maximal Abelian gauge $n=1$ is ``large enough'' (values obtained with $n>1$ typically differ less than $10\%$ from the $n=1$ ones), 
for the so-called local unitary gauges it is not~\cite{ddmo,dd}; we thus use just the elementary cubes in the gauge fixing, which is 
also what is done in the literature. We note, however, that this problem should worsen as we approach the continuum limit: as $\beta\to\infty$
the physical scale grows and it is no more allowed to use $n=1$.  

Once the monopoles are identified in the maximal Abelian gauge fixed configuration, the transformation \eqref{fma} has to be 
applied, identifying the $z$ axis with the direction of the string which goes out of the monopole; we assume that the monopole 
is located in the center of the cube. It is possible for the monopole to be located inside 
a cube with more than one plaquette pierced by a string; if this happens it would be necessary to follow the strings and 
look for loops to recognize which is the one to be used as the $z$ axis. For simplicity we restrict ourselves to the sample of monopoles 
with only one string, which should also be the great majority in the continuum limit. The gauge transformation \eqref{fma} 
is thus applied to all the links of the elementary cube enclosing the monopole; the Abelian links are again extracted and 
the new magnetic charge inside the cube calculated. This charge obviously assumes only discrete values, so that to verify 
the relation \eqref{qratio} we have to perform an average over the ensemble of the monopoles found in the configurations. 

Since in the entire procedure we have to operate only on the elementary cubes containing the monopoles, we can use a small 
lattice, of size $4\times 8^3$ (the results were nevertheless checked for consistency also on a $4\times 16^3$ 
lattice). To compute mean values, $3\times 10^4$ independent gauge configurations were generated at three different $\beta$ 
values: one below the deconfinement transition ($\beta=2.2$), and two above the transition ($\beta=2.5$ and $\beta=2.9$).

\begin{figure}[ht]
\vspace{1mm}
\scalebox{0.3}{\rotatebox{-90}{\includegraphics{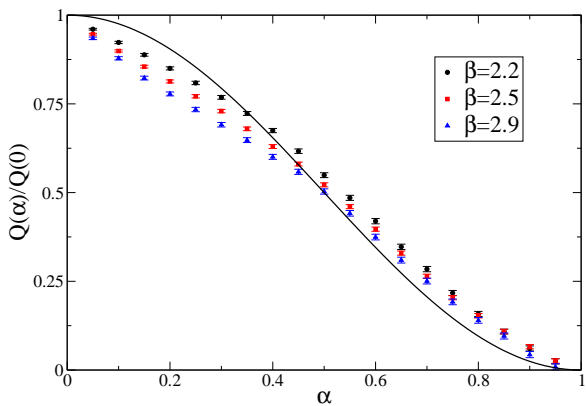}}}
\caption{Results and theoretical prediction (black line) for the ratio defined in \eqref{qratio}.}
\label{ris_fig}
\end{figure}

The results of simulations are shown in Fig.~\eqref{ris_fig} together with the theoretical prediction Eq.~\eqref{qratio}. We
have to note that several sources of systematic error are present:
\begin{enumerate}
\item we supposed the monopole to be in the center of the cube;
\item the direction of the string is identified with an angular precision of only $\pi/4$;
\item Eq.\eqref{mag} is not exactly satisfied;
\item the gauge field is defined only on the links and not pointwise.
\end{enumerate}
We have tested that the result is not much affected by the sources of error (1) and (2). The consequences of the lattice 
discretization (3) on the argument of the previous section are hard to quantify. Nevertheless numerical results are in 
good qualitative agreement with the theoretical expectations: for all nonvanishing $\alpha$ values the number of observed monopoles is 
strictly lower than the one in the maximal Abelian gauge and it approaches zero as $\alpha$ goes to $1$ (Landau gauge).

The deviations from the expected behavior and the residual $\beta$ dependence in Fig.~\ref{ris_fig} are probably due to lattice artifacts. 
Understanding in detail these artifacts goes beyond the scope of the present paper, whose main aim was to demonstrate that the gauge 
invariance of monopoles is not in contradiction with the gauge dependence of the number of monopoles observed in lattice configurations.

\section{Conclusions}

In recent years evidence has been obtained that the existence of flux tubes is a feature not only of
the maximal Abelian gauge, but also other gauges, such as local unitary gauges, random Abelian 
projection gauge, and Landau gauge (\cite{sco,scol,suzu, dmo}). These observations, together with the ones obtained by 
looking for monopole condensation in Refs.~\cite{dlmp,ccos,3}, strongly support the dual superconductor picture of gauge 
theories vacuum and the result of \cite{bdlp} that monopole condensation is a gauge-invariant phenomenon.

In this work we have analyzed some of the consequences of the results derived in \cite{bdlp} for the numerical detection of
monopoles in lattice configurations. The aim was to demonstrate the gauge invariance of the monopoles, an obvious 
prerequisite for any mechanism of confinement based on dual superconductivity.
The use of non-Abelian Bianchi identities (see \cite{bdlp}) allows us to define monopoles in a gauge-invariant way and to
explain why the observation of monopoles in lattice configurations does instead depend on the gauge.
 
To test these results we considered a class of gauges interpolating between the maximal Abelian and the Landau gauge and
we calculated by a semiclassical model the expected variation in the number of the monopoles. Numerical results qualitatively agree with
this theoretical prediction.
In particular we gave the first proof that in Landau gauge no monopoles can be detected by using the recipe of 
Ref.~\cite{dgt}, a long fact observed in numerical simulations but never understood.

\end{document}